\documentstyle[floats,aps,epsf]{revtex}
\begin{document}
\twocolumn[\hsize\textwidth\columnwidth\hsize\csname@twocolumnfalse%
\endcsname
\title{Coupled Fluctuations near Critical Wetting}
\author{A.\ O.\ Parry, C.\ J.\ Boulter and P.\ S.\ Swain}
\address{Department of Mathematics, Imperial College, London SW7 2BZ. }
\draft
\maketitle
\begin{abstract}
Recent work on the complete wetting transition has emphasized the role played by the coupling of fluctuations of the order parameter at the wall and at the depinning fluid interface. Extending this approach to the wetting transition itself we predict a novel crossover effect associated with the decoupling of fluctuations as the temperature is lowered towards the transition temperature $T_W$. Using this we are able to reanalyse recent Monte-Carlo simulation studies and extract a value $\omega(T_W) \simeq 0.8$ at ${T_W \over T_C} \simeq 0.9$ in very good agreement with long standing theoretical predictions.
\end{abstract}
\vskip2pc]

A long standing controversy in the study of phase transitions at interfaces concerns the nature of the continuous wetting transition in three dimensional systems with short ranged forces which corresponds to the marginal dimensionality \cite{brezin,dfisher,fisher1,evans,binder1,parry1,fisher2}. Renormalization group (RG) analyses of simple capillary wave models predict \cite{brezin,dfisher} strong non-universality for critical exponents and amplitudes dependent on the value of the `so called' wetting parameter $\omega(T)$ at the transition temperature $T_W$. However reliable estimates for $\omega(T)$ appropriate to the Ising model are significantly larger than values fitted to extensive Monte-Carlo simulation data \cite{binder1} which only reveal small deviations from mean field (MF) theory \cite{parry1}. Recently it has been suggested \cite{fisher2} that the transition in the Ising model is actually fluctuation induced (weakly) first-order hinting that the fitted values \cite{binder1,parry1} are unreliable. However no quantitative analysis of the simulation data was given and the question, what value of $\omega(T)$ is consistent with simulation studies of the Ising model remains unanswered.

In this paper we extend our recent analysis of fluctuation effects at the complete wetting transition based on a `two field' effective Hamiltonian $H_2[l_1,l_2]$ \cite{boulter1,parry2,boulter2} to the problem of the wetting transition discussed above. This approach predicts a novel crossover effect as the temperature is lowered towards $T_W$ which is not predicted by simple capillary wave models \cite{brezin,dfisher} (or their extensions incorporating a position dependent stiffness coefficient \cite{fisher2}) and is associated with the decoupling of fluctuations at the wall and at the fluid interface. This behaviour is in qualitative agreement with recent Monte-Carlo simulation studies of finite size (FS) effects in thin magnetic films \cite{binder2} which have already been shown to be consistent with the two field theory deep in the complete wetting regime \cite{boulter1,parry2,boulter2}. Using this prediction we are able to extract a value for the capillary parameter $\omega(T)$ at (or very close to) the transition temperature $T_W$ whilst avoiding the issue of whether the transition is weakly first or second-order. We find $\omega(T_W) \simeq 0.8$ at ${T_W \over T_C} \simeq 0.9$ which is in very good agreement with the most recent series expansion estimate at this temperature \cite{fisher1}. This is close to the universal critical value of $\omega$ \cite{moldover} which has been the long standing expectation for the magnitude of the capillary parameter appropriate to the Ising model \cite{brezin}.

To begin we make some remarks about the simulation studies used to extract this value of $\omega$. Binder, Landau and Ferrenberg (BLF) \cite{binder2} consider a thin Ising film (thickness $D$ lattice spacings) with competing surface fields $H_1=-H_D$. For this geometry there are a number of theoretical predictions for the way in which lengthscales associated with wetting phenomena determine the nature of phase coexistence and criticality in the confined system \cite{parry3,parry4}. These are all confirmed by the BLF simulation studies. In particular BLF observe the predicted symmetry broken phase (in which wetting films are adsorbed at each wall) for temperatures $T<T_C(D)$ (with $T_C(D)<T_W$ \cite{parry3}) and a `soft mode' phase \cite{parry3} in the temperature window $T_C>T>T_W$ (with $T_C$ the bulk critical temperature) where an upspin-downspin interface sits on average at the centre of the thin film and whose fluctuations are controlled by an exponentially large correlation length \cite{boulter1,parry2,boulter2,parry3,parry4} 
\begin{equation}
\xi_\parallel \sim \exp({\kappa D \over 4 \theta})
\end{equation}
Here $\kappa$ is the inverse bulk true correlation length and $\theta = \theta(T,H_1)$ is a non-universal critical amplitude, the temperature dependence of which plays a crucial role in our analysis. Thus the qualitative features of both theory and simulation are in good agreement.

The same is not true of the earlier simulations of Binder, Landau and co-workers \cite{binder1} which studied wetting in a thin Ising film with {\it equal} surface fields $H_1=H_D$. This is the standard geometry modelling fluid adsorption between parallel plates where interest usually focuses on the phenomenon of capillary condensation corresponding to the coexistence of gas-like (downspin) and liquid-like (upspin) states at a FS shifted value of the bulk ordering field $H$. The observation of phases with wetting layers adsorbed at each wall for zero bulk field $H=0$ has been previously interpreted as corresponding to meta-stable states since the equilibrium must be a liquid-like capillary condensate. Such non-equilibrium states are observed above and below $T_W$. Indeed measurement of the wall susceptibility $\chi_1$ over this region forms the basis for  locating $T_W$ \cite{binder1} and the method of extracting the value $\omega_{\rm fit} \simeq 0.3$ \cite{parry1} which recall is much smaller than expected. This method of analysing the simulation data assumes that the behaviour of $\chi_1$ (and other response functions) in the parallel plate geometry is similar to the thermodynamic semi-infinite limit. However this is {\it not} the case. We have considered the MF Landau type model for adsorption in the parallel plate geometry with identical surface fields \cite{parry3,parry4} and found that the wetting {\it spinodal} temperature $T_W^{SP}$, which marks the highest temperature (in zero bulk field) for which it is possible to have a meta-stable state with adsorbed wetting films, lies very close to and below $T_W$. Thus theory and simulation for this {\it FS geometry} are not even in qualitative agreement. It is therefore not surprising that the detailed quantative behaviour of the response functions do not yield the desired critical exponents. As mentioned above these problems do not arise for the FS geometry with reversed surface fields where the MF model and simulation results are in good qualitative agreement. For this case we can be fairly sure that the simulations are probing equilibrium behaviour which we hope to quantitatively understand using effective Hamiltonian theory.

To continue we recall some pertinent ideas in the development of effective Hamiltonian models of wetting transitions. The standard capillary wave model $H[l({\bf y})]$ has the form \cite{brezin,dfisher}
\begin{equation}
H[l({\bf y})] = \int d{\bf y} \left[ {\Sigma_{\alpha \beta}(T) \over 2} (\nabla l)^2 + W(l({\bf y})) \right]  \label{1}
\end{equation}
where $\Sigma_{\alpha \beta}(T)$ is the stiffness coefficient of the fluid interface (seperating bulk $\alpha$ and $\beta$ phases) which unbinds from the wall and whose position is described by the collective coordinate $l({\bf y})$. For systems with short ranged forces the binding potential $W(l)$ is taken to have the form \cite{brezin,dfisher}
\begin{equation}
W(l) = \overline{h} l + a {\rm e}^{- \kappa l} + b {\rm e}^{-2 \kappa l} \label{2}
\end{equation}
where $\kappa$ as mentioned above is the inverse bulk correlation length of the adsorbed ($\beta$) phase \cite{fisher1,evans} and $\overline{h}$ is proportional to the bulk ordering field. At mean field level {\it critical wetting} occurs at $\overline{h} = a(T_W^{MF}) = 0 $ provided $b>0$. Similarly the {\it complete wetting} transition occurs in the limit of vanishing bulk field $\overline{h} \rightarrow 0^+$ for $T_C>T>T_W^{MF}$ where $a(T)>0$. RG theory based on (\ref{1}) and (\ref{2}) predicts critical exponents and amplitudes which are sensitive to the wetting parameter 
\begin{equation}
\omega(T) = {k_B T\kappa^2 \over 4 \pi \Sigma_{\alpha \beta}}
\end{equation}
($k_B$ being Boltzmann's constant). For values $\omega<2$ the phase boundary for critical wetting remains $a(T_W) =0$ i.\ e.\ $T_W = T_W^{MF}$ but the critical exponents are very different from MF theory. For example along the isobar $\overline{h}=0^+$ the transverse correlation length diverges (as $T \rightarrow T_W^-$) with an exponent $\nu_\parallel = (\sqrt{2} - \sqrt{\omega})^{-2}$ for ${1 \over2}<\omega<2$ \cite{brezin,dfisher}. The implications for complete wetting are less dramatic --- only critical amplitudes are sensitive to $\omega$ whilst exponents retain their MF values \cite{dfisher}.

The discrepancy between these predictions and the older Monte-Carlo simulations \cite{binder1,parry1} led Fisher and Jin \cite{fisher2,fisher3} to reassess previous derivations of $H[l({\bf y})]$ concluding that the stiffness coefficient should be replaced with a position dependent term
\begin{equation}
\Sigma_{FJ}(l;T,\cdots) = \Sigma_{\alpha \beta}(T) + \overline{a} {\rm e}^{- \kappa l} - q l {\rm e}^{-2 \kappa l} + \cdots   \label{3}
\end{equation}
although the binding potential $W(l)$ is essentially correct. The important term in (\ref{3}) is the next to leading order exponential decay which is negative ($q>0$) at the MF phase boundary $ \overline{h} = a(T_W^{MF})=0$. The coefficient $\overline{a}$ is proportional to $a$. When this position dependence is taken into account in RG calculations the wetting transition is driven first order for sufficiently small values of $\omega<\omega^*$ where the tricritical value $\omega^*$ is expected to be of order unity. Fisher and Jin estimate that the transition appropriate to the Ising model is very weakly first order (the correlation length is enormous at the transition) and that $T_W$ is very close to $T_W^{MF}$. 

The concept of a position dependent stiffness coefficient was forwarded independently by Parry and Evans \cite{parry5} who pointed out that Hamiltonians of the form (\ref{1}) could not describe next to leading order singularites of correlation functions at the complete wetting transition (which are known to exist from exact statistical mechanical sum rules). Unfortunately the position dependence of $\Sigma(l)$ explicitly derived by Fisher and Jin using crossing and integral criteria is not of the type required by Parry and Evans \cite{parry5} to satisfy full thermodynamic consistency. One way around this is to introduce a two field effective Hamiltonian $H_2[l_1,l_2]$ \cite{boulter1,parry2,boulter2}
\begin{eqnarray}
H_2[l_1,l_2]  & = & \int d{\bf y} [ {1 \over 2} \Sigma_{\mu \nu}(l_1,l_2) \nabla l_\mu .\ \nabla l_\nu \nonumber \\
& & + U(l_1)  + W_{(2)}(l_2-l_1) ]    \label{4}
\end{eqnarray}
which models the coupling of fluctuations at the wall and $\alpha \beta$ interface. The Hamiltonian (\ref{4}) may be derived from an underlying `microscopic' Landau-Ginzburg-Wilson functional $H_{LGW}[m({\bf r})]$ using a double crossing criteria in which $l_1({\bf y})$ and $l_2({\bf y})$ are collective coordinates denoting surfaces of fixed magnetization $m_1^X$ and $m_2^X$ respectively. In the approach to the complete wetting transition the collective coordinate $l_2$ unbinds from the wall whilst $l_1$ remains bound. The position dependence of the stiffness elements $\Sigma_{\mu \nu}(l_1,l_2)$ provide a very elegant explanation of the correlation function singularites which single field Hamiltonians fail to describe. In particular it is the off diagonal elements $\Sigma_{12}(l_{21}) \sim a \kappa l_{21} {\rm e}^{-\kappa l_{21}}$ (with $l_{21}=l_2-l_1$) which provide the dominant exponential decay essential for thermodynamic consistency. The term $\Sigma_{11}$ is essentially position independent and may be identified with the stiffness of the wall-$\beta$ interface $\Sigma_{w \beta}$ whilst $\Sigma_{22}(l_{21})$ is very similar to the Fisher-Jin stiffness (\ref{3}). The presence of coupled fluctuations has a rather profound effect on the critical behaviour at complete wetting. RG calculations \cite{boulter1,boulter2} predict that critical amplitudes are no longer determined by $\omega$ but by the renormalized quantity 
\begin{equation}
\overline{\omega} = \omega + {\omega_{\beta} \over 1 + (\Lambda_1 \xi_{w \beta})^{-2} }   \label{5}
\end{equation}
where $\omega_\beta = {k_B T \kappa^2 \over 4 \pi \Sigma_{w \beta} }$ and $\xi_{w \beta}$ is the (finite) correlation length at the $w\beta$ interface which may be related to $\Sigma_{w \beta}$ and the curvature of $U(l_1)$. $\Lambda_1$ is the momentum cut-off for the bound surface $l_1$. The distinction between momentum cut-offs for the surfaces of fixed magnetization $m_1^X$ amd $m_2^X$ was not addressed in our earlier discussion of coupling effects deep in the complete wetting regime. This will play an important role in our treatment of the crossover to critical wetting.

Consider for example the effective Hamiltonian (\ref{1}) with cut-off $\Lambda$. In standard interpretation the range of wavevectors allowed in the Fourier decomposition of $l({\bf y})$ is $0 \leq Q < \Lambda$ where $\Lambda \ll \sqrt{{ \Sigma_{\alpha \beta} \over k_B T}}$ corresponding to lengthscales much greater than the bulk correlation length \cite{rowlinson}. The same idea applies to the two field Hamiltonian. In fact it is easy to establish \cite{parry6} that the local cut-off $\Lambda_1$ must satisfy $\Lambda_1 \ll \sqrt{{ \Sigma_{w \beta} \over k_B T}}$ otherwise this picture of fluctuations breaks down. This is a weak inequality and we expect $\Lambda_1$ to vanish faster than $\sqrt{\Sigma_{w\beta}}$. Of course it does not mean that fluctuations of the order parameter $m({\bf r})$ with wavevectors greater than $\Lambda_1$ do not exist rather that only for sufficiently small wavevectors are they well described by an interfacial-like collective coordinate $l_1({\bf y})$ which couples to $l_2({\bf y})$. Fluctuations in $m({\bf r})$ with wavevectors greater than $\Lambda_1$ are not interfacial-like and have to be added to the model in some other way \cite{parry6}. Coupling between these modes and $l_2({\bf y})$ does not lead to renormalization of the wetting parameter.

We now apply these ideas to the wetting transition starting in the complete wetting regime and ask how the critical amplitudes describing the transition vary as the temperature is  lowered towards $T_W^{MF}$. The important features of the Hamiltonian in this limit are 

(i) the cancellation of the leading order exponential decay in $W_2(l_{21})$ and $\Sigma_{22}(l_{21})$ similar to that indicated in (\ref{2}) and (\ref{3}). Such behaviour could have been anticipated from the simpler capillary wave model and the Fisher-Jin theory.

(ii) the vanishing of the local stiffness $\Sigma_{w \beta} \sim a^2$ and hence the momentum cut-off $\Lambda_1$. These are new effects specific to the two field model.

The important behaviour related to (ii) corresponds to a decoupling of fluctuations in the order parameter $m({\bf r})$ at the wall and $\alpha \beta$ interface. Within our theory all these features, (i) and (ii), are associated with the flattening of the magnetization profile near the wall as $T \rightarrow T_W^{MF}$. Whilst we were initially worried that this was an artifact of our model inspection of the simulation results for the magnetization \cite{binder2} appears to confirm this very close to the observed $T_W$. Exactly at $T=T_W^{MF}$ the two field model is essentially identical to that of Fisher and Jin so repeating their argument \cite{fisher2} we predict that the wetting transition is fluctuation-induced (very weakly) first-order provided $\omega$ is not too big. However because the tricritical value $\omega^*$ is not known very accurately the transition in the Ising model may still be second-order. Even if the transition was first-order it is very unlikely that this could be seen directly in simulations \cite{boulter2}.

The new prediction of this analysis is a novel crossover effect associated with the decoupling of fluctuations. Deep in the complete wetting regime the coupling of fluctuations described by (\ref{4}) increases the effective value of the capillary parameter but this effect vanishes as the temperature is lowered to the wetting temperature (or more accurately $T_W^{MF}$). This has important implications for the temperature dependence of critical amplitudes.

In Fig.\ 1 we plot the values of the critical amplitude $\theta$ taken from the susceptibility measurements of BLF. At high temperatures $T \gg T_W$ the observed value of the critical amplitude is bigger than the capillary wave prediction $\theta_{CW} = 1 + {\omega \over2}$ indicative that $\omega$ should be replaced with $\overline{\omega}$ as predicted by the two field theory \cite{boulter1,boulter2}. Recall the value of $\theta$ at ${T \over T_C} \simeq 0.96$ is consistent with the estimate based on (\ref{5}) which predicts an increment $\Delta \theta \simeq 0.3$ \cite{boulter1,boulter2}. As the temperature is reduced there is clear evidence that $\theta$ decreases consistent with the prediction discussed here. The extrapolated value
\begin{equation}
\theta^+(T) = \: \stackrel{\scriptstyle \lim}{\scriptscriptstyle T \rightarrow T_W^+(H_1)} \theta(T,H_1)
\end{equation}
should be equal to $1 + {\omega \over 2}$ (if the transition is second order) and very close to this if it is (fluctuation induced) weakly first order. Using linear and cubic fits to extrapolate to the wetting temperature ${T_W \over T_C} \simeq 0.9$ we obtain $\theta^+ \sim 1.4$ which implies that $\omega(T) \sim 0.8$ at this temperature. This is very close to the series expansion result of Fisher and Wen \cite{fisher1}. Importantly there is no indication that the `old' fitted values $\omega_{\rm fit} \simeq 0.3$ \cite{parry1} and $\omega_{\rm fit} \sim 0$ \cite{binder1} are appropriate. Such values of the capillary parameter would imply $\theta^+ \sim 1.15$ or $\theta^+ \sim 1$ (corresponding to MF theory) which are totally inconsistent with the new BLF data.

Four points that are worth emphasizing are

(a) The method avoids the issue of whether the transition is first or second-order. Even if the transition is fluctuation induced first-order the wetting temperature $T_W$ is very close to $T_W^{MF}$ and coupling effects may be ignored when $\theta$ is extrapolated to $T_W$.

(b) We have carefully checked that the simulation and theory on the nature of FS effects in the parallel plate geometry are in qualitative agreement. We are confident that the simulations are probing equilibrium behaviour.

(c) The data used to extract the critical amplitude $\theta$ are taken from the susceptibility measurements of BLF which are local to the fluctuating interface. Specifically they involve the susceptibilites $\chi_n$ and $\chi_{nn}$ centred at the middle of the thin film geometry where the interface lies on average. Older simulations \cite{binder1} had relied on measurements of the wall susceptibilites $\chi_1$ and $\chi_{11}$ which require assumption of scaling methods to analyse.

(d) The correlation length $\xi_\parallel$ in the soft mode phase is very large and easily satisfies the Ginzburg criteria \cite{halpinhealy}.

\begin{figure}
\epsfxsize=7cm
\vspace{2.5cm}
\epsffile{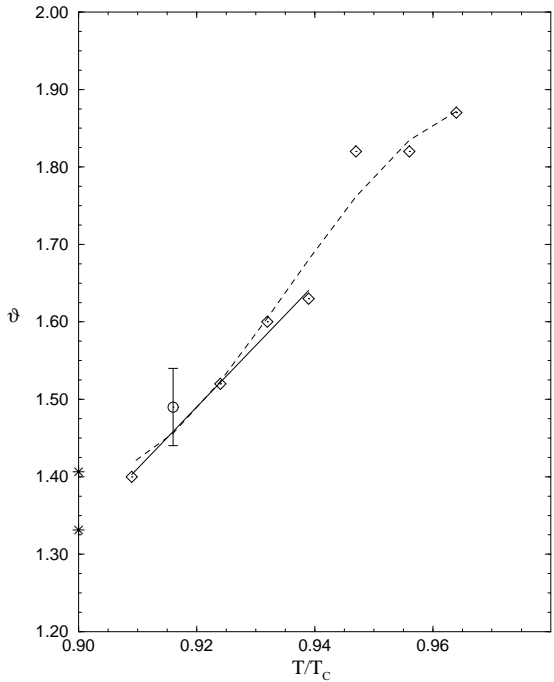}
\caption{A cubic fit of the simulation results for $\theta(T,\cdots)$. A linear fit to the lower four diamonds is also shown. The extrapolated values $\theta^+$ at ${T_W \over T_C} \sim 0.9$ are indicated by stars. Apart from the data point at ${T \over T_C} = 0.916$ the error is within the symbol. As the temperature rises from $T_W$ the value of $\theta$ increases.}
\end{figure}

In summary we have used a two field Hamiltonian model to predict a novel decoupling effect at the three dimensional wetting transition which has implications for the temperature dependence of critical amplitudes. On the basis of this we have reanalyzed recent Monte-Carlo simulations and extracted a value of the wetting parameter which is very close to long standing theoretical predictions.

This work was supported by the E.\ P.\ S.\ R.\ C.

\end{document}